\documentclass[journal,draftclsnofoot,onecolumn,12pt]{IEEEtran}

\usepackage[update,prepend]{epstopdf}

\usepackage{graphics}
\usepackage{multirow}
\usepackage{tikz}
\usepackage{bbm} 
\usepackage{pdfpages}
\usepackage{multirow}
\usepackage{subfig}
\usepackage{comment}

\usepackage{setspace}	
\usepackage{graphicx}
\usepackage{algorithm,algorithmic}
\usepackage{multicol}

\usepackage[justification=centering]{caption}
\usepackage{textcomp}
\usepackage{psfrag}
\usepackage{arydshln}
\usepackage{url}
\usepackage{soul}
\usepackage{graphicx,color}
\usepackage[nolist]{acronym}

\usepackage{mathtools,lipsum}
\usepackage{cuted}
\usepackage{amsmath}
\usepackage{graphicx}

\def\nb0{{\mathbf{0}}}
\def\nb1{{\mathbf{1}}}

\begin{document}
\graphicspath{{./Figures/}}
\newcommand{\SAR} {\mathrm{SAR}}
\newcommand{\WBSAR} {\mathrm{SAR}_{\mathsf{WB}}}
\newcommand{\gSAR} {\mathrm{SAR}_{10\si{\gram}}}
\newcommand{\Sab} {S_{\mathsf{ab}}}
\newcommand{\Eavg} {E_{\mathsf{avg}}}
\newcommand{\ft}{f_{\textsf{th}}}
\newcommand{\alphatf}{\alpha_{24}}

\title{Exploration and Application of AI in 6G Field}

\author{
Renhao Xue, Jialei Tan and Yutao Shi
\thanks{Every author has the equal contribution in this work. The authors are with University of Electronic Science \& Technology of China, Qingshuihe Campus, No. 2006, Xiyuan Ave., West High-tech Zone, Chengdu, Sichuan, China. (e-mail: 3498887948@qq.com; 1351346821@qq.com; 1721915602@qq.com)
}
\vspace{-8mm}
}

\maketitle
\begin{abstract}
The recent upsurge of diversified mobile applications, especially those supported by AI, is spurring heated discussions on the future evolution of wireless communications. While 5G is being deployed around the world, efforts from industry and academia have started to look beyond 5G and conceptualize 6G. We envision 6G to experience an unprecedented transformation that will make it completely different from the previous generations of wireless systems. In particular, 6G will go beyond mobile Internet and will be required to support AI services. Meanwhile, AI will play a critical role in designing and optimizing 6G architectures, protocols and operations. In this article, we discuss the features of 6G, and the difficulties of carrying out 6G, and AI-enabled methods for 6G network design and optimization.
\end{abstract}
\def\IEEEkeywordsname{Keywords}

\section{Introduction}
\label{sec1}

In the 1980s, 1G analog wireless cellular network can realize voice mobile communication. 10 years later, 2G cellular network based on digitalization can provide encryption services and data services, such as SMS (short message service) services. Entering the 21st century, 3G, represented by wideband code-division multiple access (WCDMA), CDMA2000, time-division synchronous CDMA (TD-SCDMA) and Worldwide Interoperability for Microwave Access (WiMAX), enabled various data services, including internet access, video calls and mobile tele-vision, after that, in the 4G/Long-Term Evolution (LTE) network started in 2009, multiple-input and multiple-output (MIMO) antenna architecture, orthogonal frequency-division multiplexing (OFDM) and all-internet protocol (IP) technology were jointly applied to achieve high-speed mobile data transmission, this makes 4G a great success \cite{zhang2022doa2}.

In recent years, we have entered the 5G era. The key technologies are network intensification, millimeter wave transmission and large-scale MIMO structure. For different application scenarios, a complete 5G communication network provides three service options: enhanced mobile broadband (eMBB); ultra-reliable low-latency communications (URLLC) and massive machine-type communications (mMTC).

With the development of communication technology, 5G standardization has been completed and has been put into commercial operation. Therefore, the vision and planning of 6G are already on the way. It aims to provide communication services for future needs after the 2030s. In the future, 6G communication aims to achieve extremely high data rates, ultra-low latency and ultra-reliability serving the Internet of everything and comprehensively supporting the development of ubiquitous intelligent mobile industry. According to the International Telecommunication Union, the number of mobile broadband access users in the world will reach 17.1 billion by 2030 \cite{union2015imt}. Here we provide a vision of 6G. In particular, people-oriented mobile communication is still the most important application of 6G.

A potential network architecture for 6G is shown in Fig. 1. We envision that AI will greatly enhance the situational awareness of the network operators, and enable closed-loop optimization to support the new service types as mentioned above. As such, 6G will unleash the full potential of mobile communications, computing, and control in a host of exciting applications, including smart cities, autonomous driving, UAVs, seamless virtual and augmented reality, Internet of Vehicles, space-air-ground integrated networks, and much more.
\begin{figure}[!ht]
\centering
\includegraphics[width=\columnwidth]{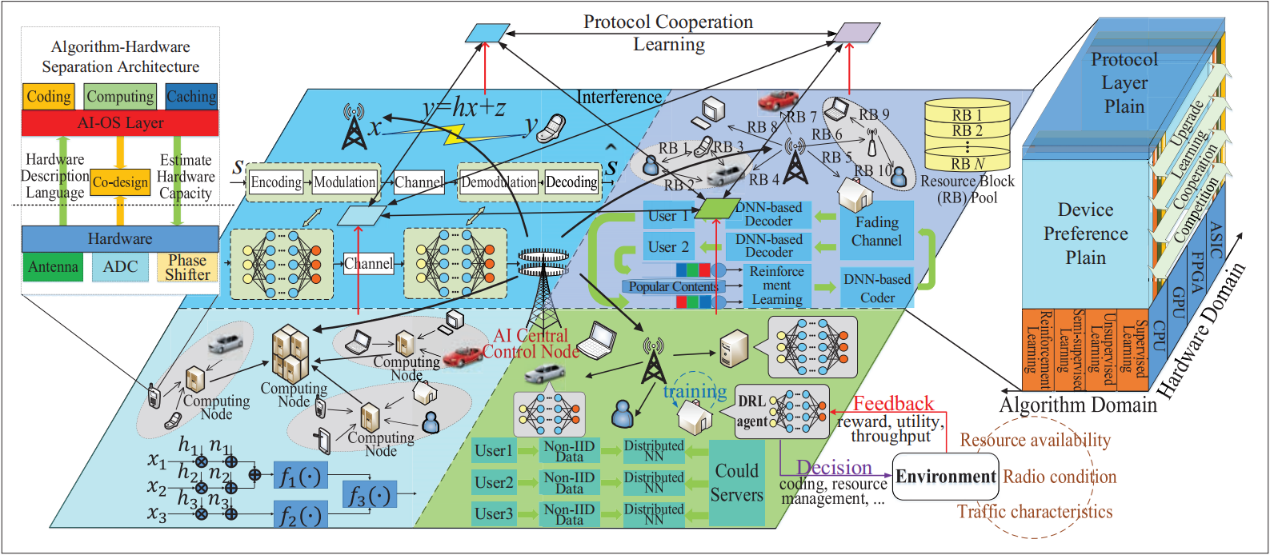}
\caption{The architecture of 6G}
\label{fig1}
\end{figure}

\section{6G Development Vision and Challenges}
\label{sec2}

\textbf{Artificial Intelligence}: In the current research on 6G, artificial intelligence is the most eye-catching idea. 6G enabled by artificial intelligence is considered to provide a series of new functions, such as self- focusing, context awareness \cite{stoica20196g}. In addition, 6G enabled by artificial intelligence will release the full potential of radio signals and realize the transformation from cognitive radio to intelligent radio. From the point of view of algorithm, machine learning is the key to the realization of 6G enabled by artificial intelligence. In addition to algorithms, the reconfigurable intelligent surface is also used to build the hardware foundation of artificial intelligence in wired communication \cite{renzo2019smart}. It is envisaged as a large-scale MIMO 2.0 in 6G. What's more, main features of 6G and enabling communication technology are also the focus of attention. Due to the limitation of Shannon limit, it is difficult to improve the spectral efficiency of 6G on a large scale. In contrast, the security, confidentiality, privacy and intelligence of 6G should be greatly improved through new technologies.

\textbf{Computation Oriented Communications (COC)}: New smart devices call for distributed computation to enable the key functionalities, such as federated learning. Instead of targeting classical quality of service (QoS) provisioning, CoC will flexibly choose an operating point in the rate-latency-reliability space depending on the availability of various communications resources to achieve a certain computational accuracy.

\textbf{Event Defined uRLLC (EDuRLLC)}: In contrast to the 5G uRLLC application scenario where redundant resources are in place to offset many uncertainties, 6G will need to support uRLLC in extreme or emergency events with spatially and temporally changing device densities, traffic patterns, and spectrum and infrastructure availability.

\textbf{Strengthen Traditional Mobile Communication}: 6G communication is people-oriented. Therefore, mobile phone is still the main communication tool. The challenges of traditional communication come from five aspects: how to enhance security and protect privacy; how to expand network coverage in a rapid and cost-efficient way, especially in distant and isolated areas; how to reduce the cost of mobile communications; how to extend the battery life of the mobile device; and how to provide a higher data rate with a lower end-to-end latency \cite{wang2022stochastic}.

\textbf{Precise Indoor Positioning}: Due to the complex indoor electromagnetic propagation environment, indoor positioning is not yet mature \cite{zhao2022wknn}. Many people believe that indoor positioning is not feasible only through RF communication, but in the 6G era, such a key and influential application is expected to be realized through more advanced non RF communication technology.

\textbf{New Communication Terminal}: It is expected that in 2030, more and more communication devices will appear, which may be wearable devices, integrated headphones or implantable sensors. These possible new communication terminals may have different system requirements and mathematical modeling. In addition, for health reasons, the transmission power and frequency band used in these devices must be strictly limited.

\begin{figure}[!ht]
	\centering
	\includegraphics[width=0.9\columnwidth]{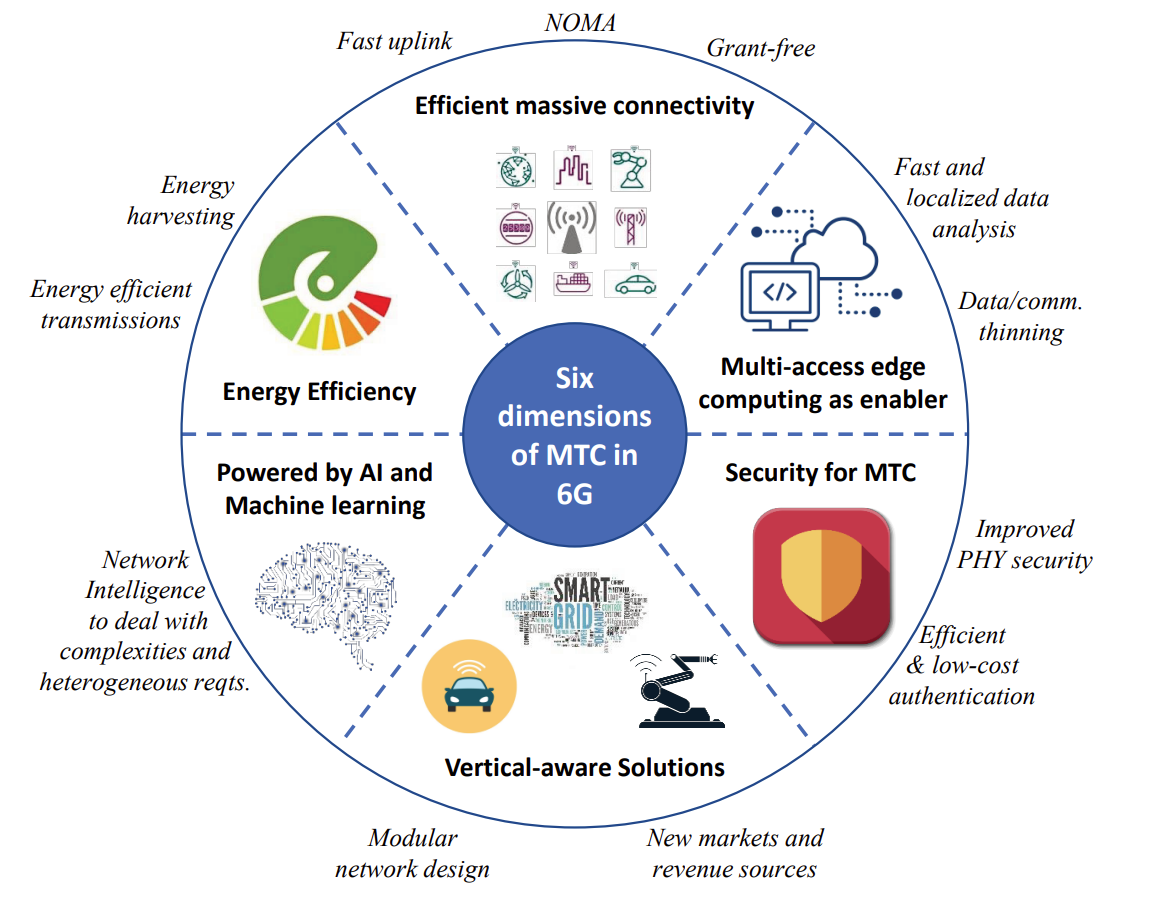}
	\caption{Six key enablers for MTC in 6G and their respective solution components}
	\label{fig2}
\end{figure}

\textbf{High Quality Communication Service on Aircraft}: In previous generations of communication, the communication task on the aircraft can not be satisfied. Its service is mainly challenged by factors such as high mobility, Doppler frequency shift, frequent handover and insufficient coverage \cite{}. In order to provide high-quality communication services on aircraft, not only new communication technologies must be adopted in 6G communication, but also new network architecture must be used \cite{wang2022ultra}.

\textbf{Global Connectivity and Integrated Networks}: In the past few years, communication research has always focused on dense urban agglomerations. It should be noted that there are still a large number of people in the world who cannot access the most basic data services, most of them are distributed in rural areas of backward developing countries \cite{belmekki2022unleashing}. The arrival of 6G era should realize the horizontal and vertical development of wireless network, benefit more people, and finally realize global connection. In order to achieve this goal, or the three-dimensional integrated network will be used, it is envisaged that in order to realize the communication network transmission of globally connected flight nodes, it will become ubiquitous in the 2030's. This should be taken into account when planning the 6G network. This three-dimensional integrated network can bring considerable performance improvement and unprecedented services to users.

\textbf{Support Vertical Industry Communications}: 6G communication is expected to continue to support applications in vertical industries. In order to integrate these industries in 6G communication, 5G MTC needs to be upgraded.

\textbf{Holographic Communication}: 6G is expected to become a conversion point from traditional video conference to virtual human conference. It aims to enable people to interact with the received holographic data and modify the received video as needed.

\textbf{Tactile Communication}: After using holographic communication to transmit virtual scenes of near real people, events, environments, etc., it is beneficial to exchange physical interactions in real time and remotely through the tactile Internet. Specifically, the expected services include remote operation, cooperative autonomous driving and interpersonal communication, in which tactile control can be applied through the communication network.

\textbf{Interpersonal Communication}: 6G is expected to widely support people-oriented communication concepts, where people can access or share physical features. Therefore, the concept of communication between people has been put forward, allowing access to the five senses of human beings. The above service types represent emerging driving applications of 6G. They can hardly be offered by 5G, not only because of their stringent requirements for higher data rates, lower latency, denser connection, and so on, but also due to their extreme demand for new performance metrics that have never been considered in 5G, for example, delay jitter, context awareness, UAV/satellite compatibility, and so on. Inspired by these trends, in this article, we attempt to conceptualize 6G as an intelligent information system that is both driven by and a driver of the modern AI technologies.

\begin{figure}[!ht]
\centering
\includegraphics[width=0.6\columnwidth]{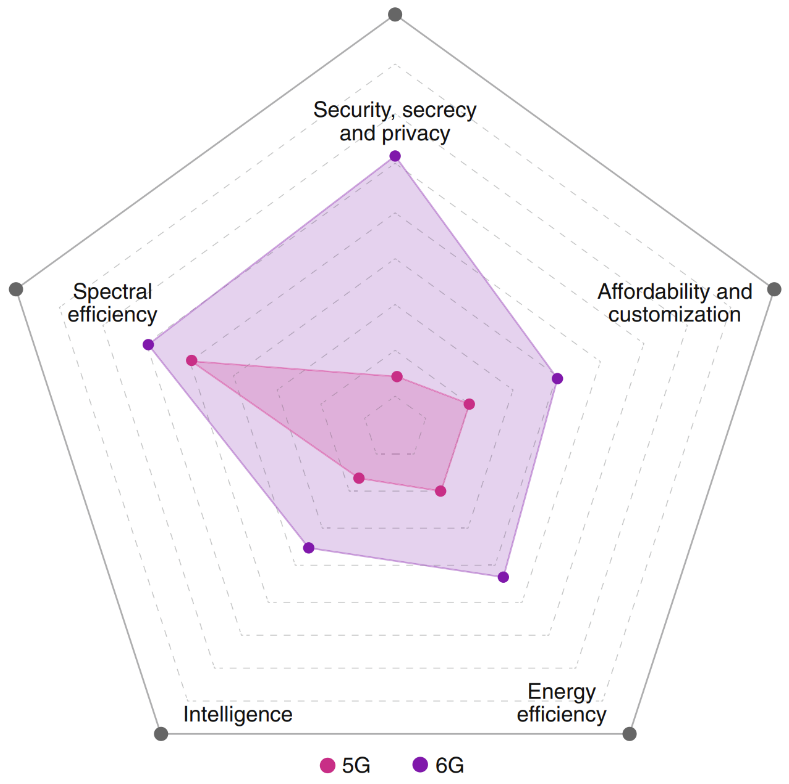}
\caption{Qualitative comparison between 5G and 6G communications}
\label{fig3}
\end{figure}

\begin{figure}[!ht]
\centering
\includegraphics[width=\columnwidth]{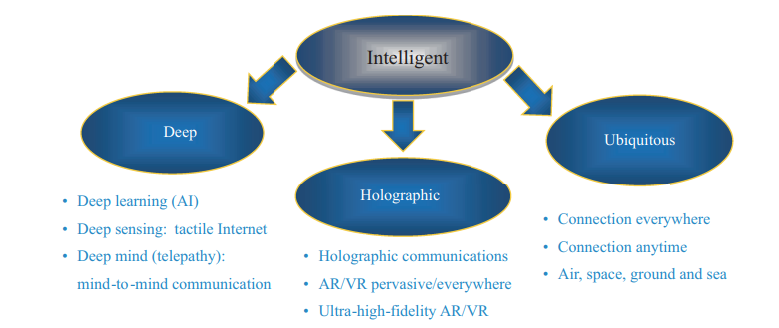}
\caption{6G vision}
\label{fig4}
\end{figure}

\section{Security and Privacy Issues in 6G}
\label{sec3}

6G envisions the realization of the Internet of Everything (IoE), a massive network of billions of different devices, including many smaller devices, such as in-body sensors. The resource-constrained IoT(The Internet of Things) devices cannot afford complicated cryptography to maintain strong security \cite{saad2019vision}, making them a primary target of the attackers. These devices can be compromised and potentially used to initiate attacks. Data collection by hyper-connected IoE to serve 6G applications raises privacy issues. Because a 6G network contains many more different small devices than a 5G network, these local networks operate as a standalone. In contrast to the well-defined local 5G networks, many stakeholders implement local 6G networks with different embedded security levels. A local 6G network with minimal security provides intruders with the opportunity to initiate attacks. And then infiltrate the networks which trust the compromised network \cite{yang2017survey}. Each device has mesh connectivity, thereby increasing the threat surface.

6G rely on AI to enable fully autonomous networks. Therefore, attacks on AI systems, especially Machine Learning (ML) systems, will affect 6G. The security threats ML Systems faces include Poisoning attacks, data injection, data manipulation, logic corruption, model evasion, model inversion, model extraction, and membership inference attacks. Moreover, attacks with quantum computers to destabilize the blockchain are also feasible. This is the point where 6G and 5G face different security issues. The current 5G standard does not deal with the security issues raised by quantum computers because it relies on traditional cryptography. As the 6G era will mark the existence of quantum computers \cite{dogra2020survey}. Current cryptographic-based security mechanisms are vulnerable to attacks based on quantum computers.

As an engineering system, it is impossible to meet all the desired functions with certain resource investment. Therefore, we must make some trade-offs:

\begin{itemize}
    \item \textbf{Privacy and Intelligence}: On the one hand, artificial intelligence algorithms need to acquire and process personal data to optimize network operation, Adjust network coefficients and provide high-quality services. On the other hand, for the sake of high intelligence, privacy will be sacrificed. One potential solution is to introduce an intermediary between end-user data and AI algorithms.
    \item \textbf{Affordability and Intelligence}: High intelligence may increase the cost of network operators and equipment manufacturers. All the increased costs will increase the burden on users. In order to achieve this trade-off, it is necessary to make a breakthrough in intelligent systems. In smart grid, the business strategy of power users selling their own power back to power companies may be borrowed by 6G network.
    \item \textbf{Customization and Intelligence}: Sometimes, users' preferences may not be completely consistent with the optimization algorithm generated by artificial intelligence algorithm. This contradiction is expressed as the trade-off between customization and intelligence in 6G communication. It should be considered that the priority of customization should always be higher, that is, artificial intelligence algorithms and intelligent devices should establish prohibitive clauses, and should be in the bottom layer protocol of 6G communication.
\end{itemize}

\begin{figure}[!ht]
	\centering
	\includegraphics[width=0.8\columnwidth]{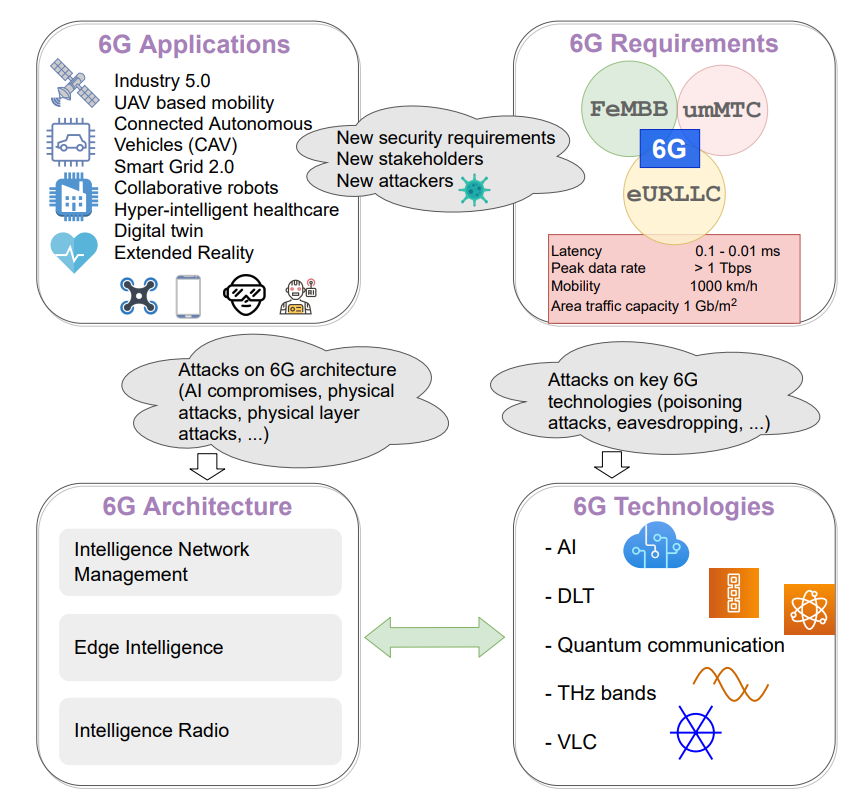}
	\caption{6G landscape and security composition}
	\label{fig5}
\end{figure}

\begin{figure}[!ht]
\centering
\includegraphics[width=0.6\columnwidth]{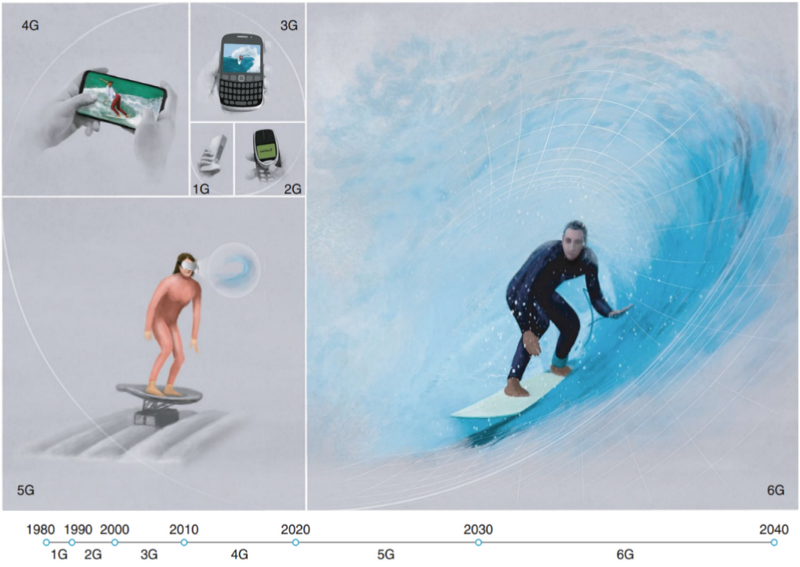}
\caption{A user’s perception of the different communications networks, from 1G to the hypothetical 6G}
\label{fig6}
\end{figure}

\section{AI Applications in 6G}
\label{sec4}

\subsection{AI for Security and Privacy in 6G}
\label{sec4.1}
In 6G network, the volume and variety of data are growing which makes it possible to apply the big data analytic. We are able to mine historical data to get insights on network performance, traffic profile, channel conditions, user perspectives, and so on. Then make operators and service providers aware of the situation of network. Also, big data analysis enable autonomous detection of network faults and service impairments, identify the root causes of network anomalies, and ultimately improve the network reliability and security. 

Another use is to predict future events such as traffic patterns, user locations, user behavior and preference, and resource availability. Last but not least, big data analysis take advantage of the predictions to suggest decision options for resource allocation, edge computing and so on. The predictive analysis using AI can predict attacks such as attacks on blockchain before the attack occurs. It must be mentioned that harvesting and analyzing a large amount of data raise concerns about data security, privacy and ownership. So, the 6G architecture and protocols should be designed in a way that protects data security, privacy and integrity.

Multi-connectivity, mesh networks with tiny cells in 6G allow simultaneous communication for devices via multiple base stations. Edge-based ML models could be used for dynamic detection of privacy-preserving routes~\cite{2}. Federated learning keeps data in the user's proximity compared to cloud-based centralized learning to enhance data privacy and location privacy.

\subsection{Closed-loop Optimization}
\label{sec4.2}
Traditional methods for wireless network optimization may not be applicable in 6G, as the network will be extremely dynamic and complex due to the scale, density, and diversity. In the complex 6G network environment, the mapping between a decision and its effect on the physical system is cost highly to define and may not be analytically available. Recent advances in AI technologies can establish a feedback loop between the decision maker and the physical system, so that the decision maker can optimize its action based on the system's feedback to reach optimality eventually.

\subsection{Intelligent Wireless Communications}
\label{sec4.3}

Unavoidably, the physical structure of wireless communication systems suffers from variety of impairments. To communicate reliably and efficiently with combinations of hardware, a large number of design parameters need to be controlled and optimized jointly. In the past, end-to-end optimization has never been practical in wireless system due to the high complexity. Instead, existing approaches divide the whole system to multiple independent blocks, each has a simplified model that does not capture the feather of real-world systems. AI technologies open up the possibilities in end-to-end optimization of the full chain of the physical layer, from the transmitter to the receiver. We envision an “intelligent physical layer” in 6G, where the end-to-end system is capable of self-learning and self-optimization by combining advanced sensing and data collection, AI technologies, and domain-specific signal processing approaches \cite{zhang2022doa}.
\begin{figure}[!ht]
\centering
\includegraphics[width=\columnwidth]{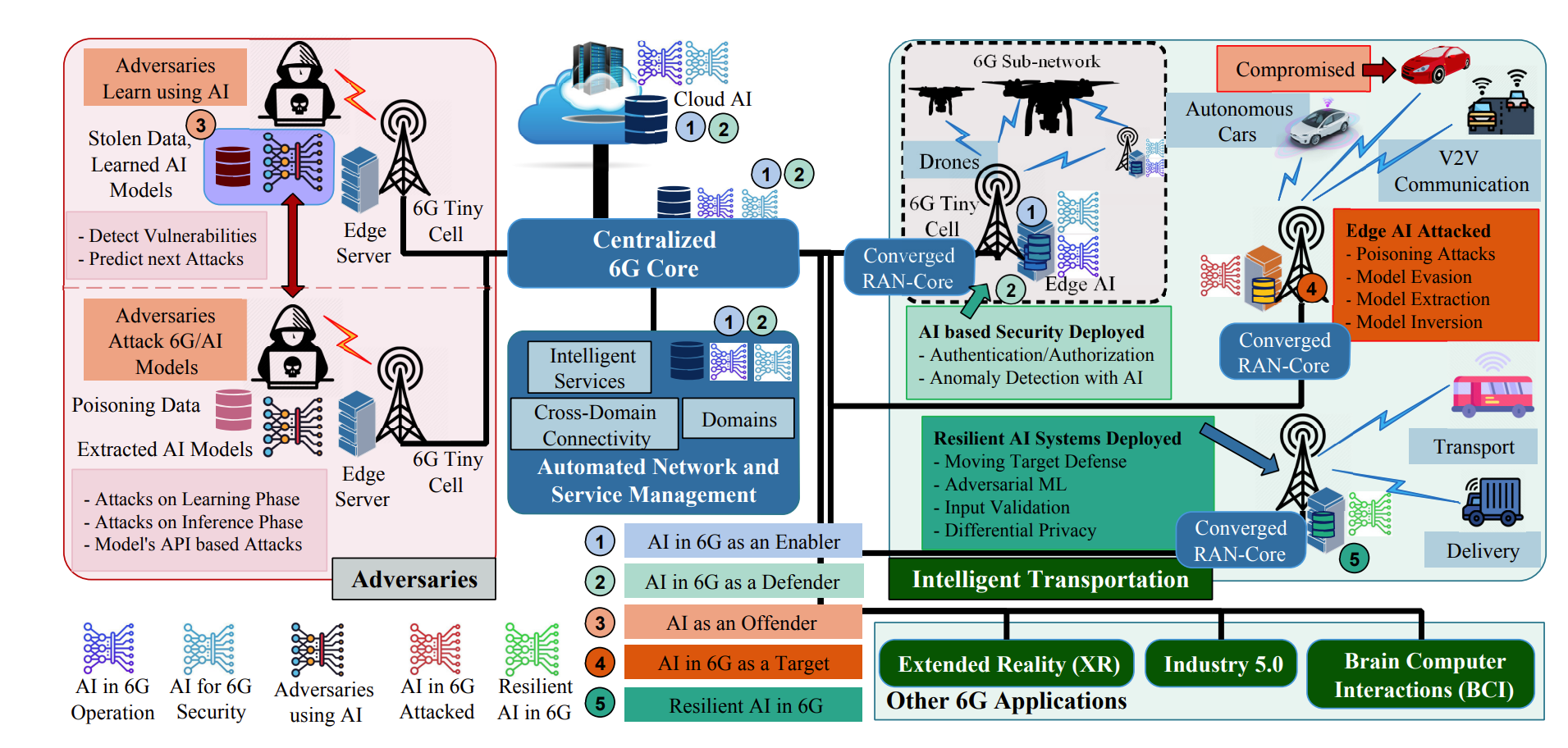}
\caption{AI’s role as an Enabler, Defender, Offender and a Target in 6G Intelligent Transportation}
\label{fig7}
\end{figure}

\section{Further Discussions beyond Communications}
\label{sec5}
Communication technologies are crucial, but not all. To promote a new technological paradigm and make it socioeconomically profitable, we must always keep the issues beyond technology in mind. In this section, we briefly discuss several crucial issues relating to 6G beyond the communication technologies themselves and hope to attract attention in the 6g development.

\textbf{Dependence on Basic Disciplines}: The development of wireless communication is highly restricted by basic science, especially mathematics and physics. The current mathematical tools make it impossible for us to explore the exact performance of communication systems. Limited by this, a large number of unrealistic assumptions can not provide accurate guidance for 6G communication.

\textbf{Dependence on Upstream Industries}: In the theoretical research, people can assume that any high frequency can be used, but in reality, the communication equipment composed of real electronic components may not meet the relevant requirements and sometimes exceed the constraints of electronic circuits. Therefore, we cannot ignore the dependence on upstream enterprises.

\textbf{Business Model and Commercialization of 6G}: Previous research activities mainly focused on the technology itself, ignoring the business model and commercialization. Ignoring these results in failure. For example, in response to emergencies, network intensification is a good choice. The question is, who will pay for the construction of expensive base stations? In addition, how to ensure the compatibility of 6G with 4G/LTE and 5G? These are questions to be discussed.

\textbf{Potential Health and Psychological Problems of Users}: The radiation problem of base stations has always aroused people's concern. With the arrival of 6G, the network will become more intensive. Perhaps this concern will become more and more serious. Considering that the theme of 6G is people-oriented, this issue should not be ignored \cite{lou2021green}.

\textbf{Future Work}: AI is a key driver for the next generation of 6G mobile networks and ensuring security is a key consideration in realizing the 6G vision. This article outlines the security and privacy issues in the current 6G development and discusses the role that AI can play in addressing these issues. In the future, we expect AI to contribute to the further improvement of 6G secure communication.

\section{Conclusion}
\label{sec6}

This paper first briefly introduces the current research progress of 6G. Then it analyzes and introduces some vision and challenges of 6G. After that, we discussed the possible security and privacy issues in 6G. And we gave the trade-offs between key features and potential solutions. In the third section, we discussed how AI can help solve some problems and some other applications of AI in 6G. Finally, we discussed and put forward other key issues besides communication technologies and security that should be considered in the development of 6G.



\bibliographystyle{IEEEtran}
\bibliography{references}

\begin{thebibliography}{10}
\providecommand{\url}[1]{#1}
\csname url@samestyle\endcsname
\providecommand{\newblock}{\relax}
\providecommand{\bibinfo}[2]{#2}
\providecommand{\BIBentrySTDinterwordspacing}{\spaceskip=0pt\relax}
\providecommand{\BIBentryALTinterwordstretchfactor}{4}
\providecommand{\BIBentryALTinterwordspacing}{\spaceskip=\fontdimen2\font plus
\BIBentryALTinterwordstretchfactor\fontdimen3\font minus
  \fontdimen4\font\relax}
\providecommand{\BIBforeignlanguage}[2]{{%
\expandafter\ifx\csname l@#1\endcsname\relax
\typeout{** WARNING: IEEEtran.bst: No hyphenation pattern has been}%
\typeout{** loaded for the language `#1'. Using the pattern for}%
\typeout{** the default language instead.}%
\else
\language=\csname l@#1\endcsname
\fi
#2}}
\providecommand{\BIBdecl}{\relax}
\BIBdecl

\bibitem{zhang2022doa2}
X.~Zhang, Z.~Zheng, W.-Q. Wang, and H.~C. So, ``{DOA} estimation of coherent
  sources using coprime array via atomic norm minimization,'' \emph{IEEE Signal
  Processing Letters}, 2022.

\bibitem{union2015imt}
I.~Union, ``{IMT} traffic estimates for the years 2020 to 2030,'' \emph{Report
  ITU}, vol. 2370, 2015.

\bibitem{stoica20196g}
R.-A. Stoica and G.~T.~F. de~Abreu, ``{6G}: the wireless communications network
  for collaborative and {AI} applications,'' \emph{arXiv preprint
  arXiv:1904.03413}, 2019.

\bibitem{renzo2019smart}
M.~D. Renzo, M.~Debbah, D.-T. Phan-Huy, A.~Zappone, M.-S. Alouini, C.~Yuen,
  V.~Sciancalepore, G.~C. Alexandropoulos, J.~Hoydis, H.~Gacanin \emph{et~al.},
  ``Smart radio environments empowered by reconfigurable ai meta-surfaces: An
  idea whose time has come,'' \emph{EURASIP Journal on Wireless Communications
  and Networking}, vol. 2019, no.~1, pp. 1--20, 2019.

\bibitem{wang2022stochastic}
R.~Wang, M.~A. Kishk, and M.-S. Alouini, ``Stochastic geometry-based low
  latency routing in massive leo satellite networks,'' \emph{arXiv preprint
  arXiv:2204.03802}, 2022.

\bibitem{zhao2022wknn}
Z.~Zhao, Z.~Lou, R.~Wang, Q.~Li, and X.~Xu, ``{I-WKNN}: Fast-speed and
  high-accuracy {WIFI} positioning for intelligent sports stadiums,''
  \emph{Computers \& Electrical Engineering}, vol.~98, p. 107619, 2022.

\bibitem{wang2022ultra}
R.~Wang, M.~A. Kishk, and M.-S. Alouini, ``Ultra-dense {LEO} satellite-based
  communication systems: {A} novel modeling technique,'' \emph{IEEE
  Communications Magazine}, vol.~60, no.~4, pp. 25--31, 2022.

\bibitem{belmekki2022unleashing}
B.~E.~Y. Belmekki and M.-S. Alouini, ``Unleashing the potential of networked
  tethered flying platforms: {P}rospects, challenges, and applications,''
  \emph{IEEE Open Journal of Vehicular Technology}, vol.~3, pp. 278--320, 2022.

\bibitem{saad2019vision}
W.~Saad, M.~Bennis, and M.~Chen, ``{A} vision of {6G} wireless systems:
  Applications, trends, technologies, and open research problems,'' \emph{IEEE
  network}, vol.~34, no.~3, pp. 134--142, 2019.

\bibitem{yang2017survey}
Y.~Yang, L.~Wu, G.~Yin, L.~Li, and H.~Zhao, ``A survey on security and privacy
  issues in {I}nternet-of-{T}hings,'' \emph{IEEE Internet of Things Journal},
  vol.~4, no.~5, pp. 1250--1258, 2017.

\bibitem{dogra2020survey}
A.~Dogra, R.~K. Jha, and S.~Jain, ``A survey on beyond {5G} network with the
  advent of {6G}: Architecture and emerging technologies,'' \emph{IEEE Access},
  vol.~9, pp. 67\,512--67\,547, 2020.

\bibitem{zhang2022doa}
X.~Zhang, Z.~Zheng, W.-Q. Wang, and H.~C. So, ``{DOA} estimation of mixed
  circular and noncircular sources using nonuniform linear array,'' \emph{IEEE
  Transactions on Aerospace and Electronic Systems}, 2022.

\bibitem{lou2021green}
Z.~Lou, A.~Elzanaty, and M.-S. Alouini, ``Green tethered {UAV}s for {EMF}-aware
  cellular networks,'' \emph{IEEE Transactions on Green Communications and
  Networking}, vol.~5, no.~4, pp. 1697--1711, 2021.

\end{thebibliography}

\end{document}